\documentclass[a4paper]{article}

\usepackage{INTERSPEECH2022,booktabs,xfrac,multirow,hyperref,cleveref}

\DeclareMathOperator*{\argmax}{arg\,max}

\title{UniKW-AT: Unified Keyword Spotting and Audio Tagging}
\name{Heinrich Dinkel$^{\ddagger}$, Yongqing Wang$^{\ddagger}$, Zhiyong Yan$^{\ddagger}$, Junbo Zhang and Yujun Wang\thanks{$^{\ddagger}$ equal contribution.}}
\address{
  Xiaomi Corperation, Beijing, China}
\email{\{dinkelheinrich,wangyongqing3,yanzhiyong,zhangjunbo1,wangyujun\}@xiaomi.com}

\begin{document}

\maketitle
\begin{abstract}
Within the audio research community and the industry, keyword spotting (KWS) and audio tagging (AT) are seen as two distinct tasks and research fields.
However, from a technical point of view, both of these tasks are identical: they predict a label (keyword in KWS, sound event in AT) for some fixed-sized input audio segment.
This work proposes UniKW-AT: An initial approach for jointly training both KWS and AT.
UniKW-AT enhances the noise-robustness for KWS, while also being able to predict specific sound events and enabling conditional wake-ups on sound events.
Our approach extends the AT pipeline with additional labels describing the presence of a keyword.
Experiments are conducted on the Google Speech Commands V1 (GSCV1) and the balanced Audioset (AS) datasets.
The proposed MobileNetV2 model achieves an accuracy of 97.53\% on the GSCV1 dataset and an mAP of 33.4 on the AS evaluation set.
Further, we show that significant noise-robustness gains can be observed on a real-world KWS dataset, greatly outperforming standard KWS approaches.
Our study shows that KWS and AT can be merged into a single framework without significant performance degradation.
\end{abstract}

\section{Introduction}

Keyword spotting (KWS) or wake word detection is nowadays a crucial front-end task for most intelligent voice assistants.
When a user utters a specific key phrase, the keyword, an assistant wakes up and provides the user some service e.g., a weather report.
Audio tagging (AT) is a task that aims to label a specific audio content into sound event classes e.g., a cat meowing or a dog barking.
One of the current applications of AT models is the assistance of hearing-impaired people, to notify a user that some event inaudible to the user has happened e.g., a door has been knocked.

From a technical standpoint of view, KWS and AT are identical in their implementation.
Specifically, given an input sample of some specific length e.g., 1 s, the goal of both tasks is to provide an output label signaling the presence of a keyword (KWS) or a sound event (AT).
However, most real-world applications for these two tasks are exclusive: Either KWS or AT can be operational at a time.

This work proposes {uni}fied keyword {a}udio {t}agging (UniKW-AT), which is capable of jointly modeling both tasks using a single model.
The potential benefits of UniKW-AT are an improved robustness of a KWS model in regards to real-world noises, such as music, and the possibility to provide sound event information to the user.
Additional benefits are the capability to incorporate different types of speech into the KWS post-processing, such as whispering or children's speech.

Nonetheless, there are inherent problems when jointly training a single model for KWS and AT.
First, training large-scale AT models utilizes weakly supervised training, since large-scale strongly labeled datasets are costly~\cite{hershey_icassp2021_strong_labels_audioset}.
On the contrary, KWS models are commonly trained with strong supervision, where the keyword position in time is known.
Second, AT models are trained with a multi-label objective, since multiple sound events can co-occur within some time segment, e.g., music playing in the background while people are talking.
On the contrary, standard KWS models are commonly trained using a single-label objective such as categorical cross-entropy.

Our core contribution in this work is the first attempt at joint KWS and AT modeling, with a focus on a noise-robust KWS performance and sound event detection capability.

\subsection{Previous work}

Most previous work regarding KWS is focused on decreasing a model's parameter size and increasing its inference speed~\cite{Mo2020}.
Works such as~\cite{sainath15b_interspeech,kim21l_interspeech,choi19_interspeech} used convolutional neural networks (CNNs) for this matter, while more recently transformer based models~\cite{berg21_interspeech, sahu2021audiomer} and multi-layer perceptron (MLP) mixers~\cite{morshed2021attention,ng2022convmixer} have also received attention from the community.

On the other hand, works for AT generally are performance-oriented, meaning that few works exist that focus on practical implications of models such as inference time or model size.
Similar to KWS research, CNNs dominate the research for AT, with notable examples being~\cite{Kong2020d,fonseca2020fsd50k}.
Further, transformers have also been recently seen success~\cite{gong21b_interspeech,chen2022hts,koutini2021efficient}.
However, previously introduced models in literature aimed at AT cannot be used for KWS, since KWS is an always-on feature computed on-device and previous work using CNN/Transformer models require a GPU for fast inference.
Thus, our work focuses on the usage of lightweight and fast models.

This paper is structured as follows.
\Cref{sec:method} establishes our proposed UniKW-AT framework.
Then \Cref{sec:experiments} introduces the experiments and \Cref{sec:results} displays the results of this work.
Lastly, \Cref{sec:discussion} discusses the shortcomings of our work and \Cref{sec:conclusion} concludes the paper.

\section{Method}
\label{sec:method}

\begin{figure}[htbp]
    \centering
    \includegraphics[width=0.98\linewidth]{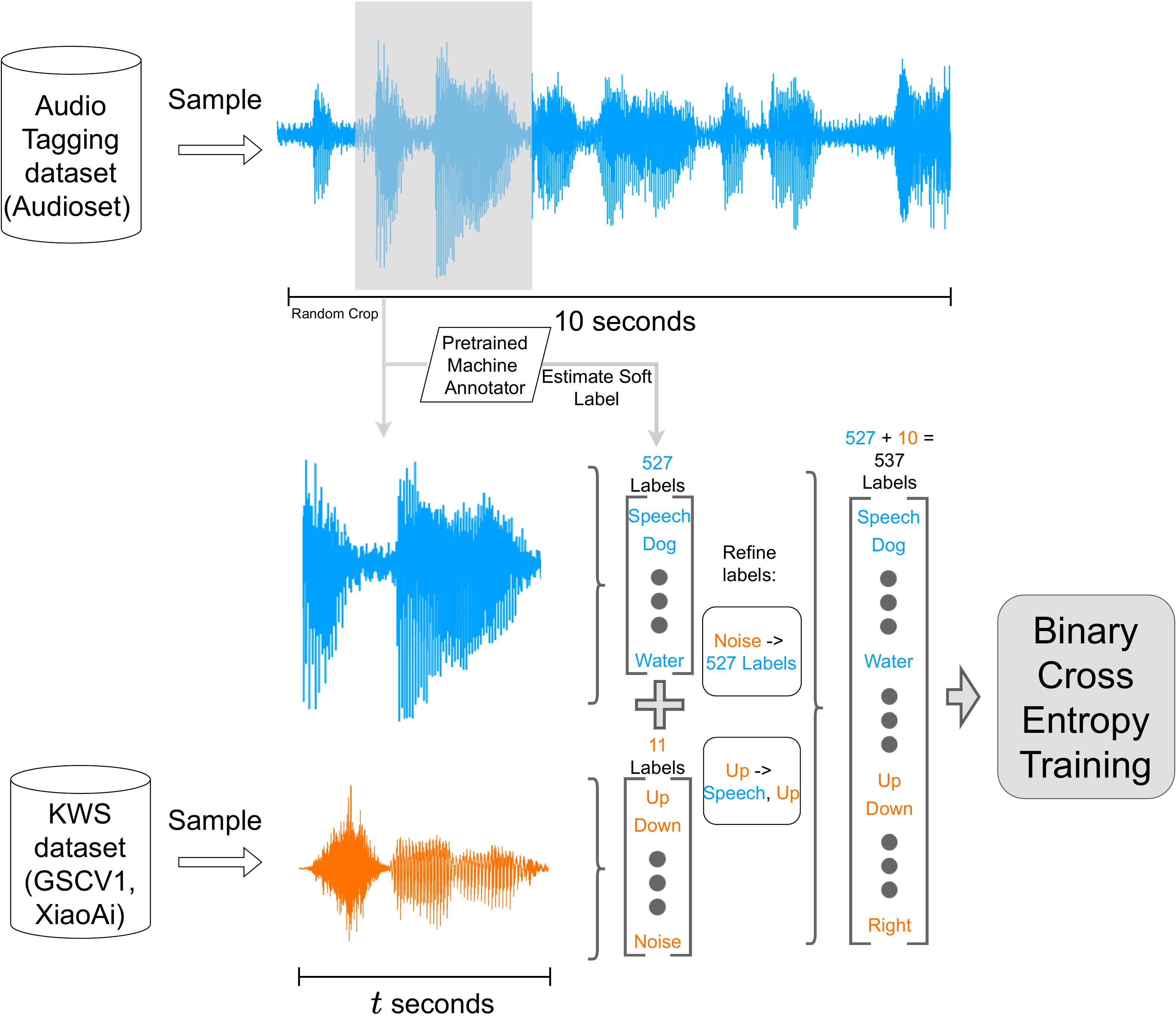}
    \caption{Depiction of the proposed UniKW-AT approach. The label sets of KWS and AT are merged and the ``noise'' label in KWS is replaced it with the ``Speech'' label in the AT branch. The AT branch (top) is trained using pseudo strong labels (PSL) by estimating soft labels from a pretrained machine annotator on random crops matching the length of the KWS dataset ($t$). Training optimizes the binary cross entropy (BCE) criterion.}
    \label{fig:framework}
\end{figure}

Our approach jointly models KWS and AT by treating KWS labels as additional target events within the standard AT training framework.
Given the target KWS labelset $\mathbb{L}_{\text{KWS}}$ with $K$ keywords and an AT labelset $\mathbb{L}_{\text{AT}}$ with $C$ sound events, we aim to model both tasks by merging both labelsets obtaining $\mathbb{L} = \mathbb{L}_{\text{KWS}} \cup \mathbb{L}_{\text{AT}}$.
Specifically, our method uses the $C=527$ sound event labels of Audioset (AS)~\cite{gemmeke2017audio} as $\mathbb{L}_{\text{AT}}$.
Models trained for KWS commonly contain a non-target label, also known as ``noise''  in their label set.
In the case of UniKW-AT, this additional ``noise'' label is not needed, since most types of noise are already encompassed within the $\mathbb{L}_{\text{AT}}$ label set.
Our approach first appends all $K$ target keywords to the $C=527$ available AS labels, resulting in $C+K$ target labels.
All non-target keywords contained within a KWS dataset are mapped to the ``Speech'' label within $\mathbb{L}_{\text{AT}}$.

Given a KWS dataset and an AT dataset, we preprocess the raw audio samples, such that the audio clip durations from both datasets match.
Samples from both datasets are randomly cropped to some target duration $t$ i.e., 1 s.
Randomly cropping the AT sample leads to a label misalignment problem, since labels are provided on a 10 s scale.
This problem is partially solved by using a pre-trained machine annotator to create pseudo strong labels (PSL)~\cite{dinkel2022_icassp} as the AT target.

We use the multi-label binary cross-entropy (BCE) objective in this work as our default training metric but provide comparisons to the standard cross-entropy (CE) objective in \Cref{ssec:noise_robustness}.
We choose BCE in favor of CE due to its practical advantages during inference.
For most real-world tasks, only a fraction of the provided $C$ sound event labels are relevant.
In the case of BCE trained networks, undesirable labels (weights in the final linear layer) can be removed, since each label's output probability is independent.
This decreases model size, increases inference speed, and reduces power consumption while retaining performance for the labels of choice.
The proposed framework is depicted in \Cref{fig:framework}.

\section{Experiments}
\label{sec:experiments}

\subsection{Datasets}

\paragraph*{Training}

We make use of two training datasets being Google Speech Commands V1 (GSCV1)~\cite{warden2018speech} and AS~\cite{gemmeke2017audio} in this work.
GSCV1 contains 65,000 utterances of 30 different keywords, i.e., ``Up'', ``Right'', ``Down'', spoken by thousands of people.
We use the common 11 class subset of GSCV1, where the original 30 classes have been reduced to 10 common keywords: ``Yes'', ``No'', ``Up'', ``Down'', ``Left'', ``Right'', ``On'', ``Off'', ``Stop'', ``Go'' while the rest 20 keywords are mapped to ``unknown/noise''.
The official training/validation/testing split containing 51,088/6,798/6,835 utterances, respectively, is used.
Each sample within GSCV1 is at most 1 s long.

As for AT training, we use the 60 h long balanced subset of AS containing 21,292 audio clips with a duration of at most 10 s per clip.

\paragraph*{Evaluation}

Evaluation is split between the KWS and AT subtasks.
AT evaluation uses the common evaluation subset of AS, containing 18,229 audio clips with an overall duration of 50 h.

KWS evaluation mainly focuses on the GSCV1 dataset, where 2,567 target keywords and 4,268 non-target samples are provided.
Further, to explore the robustness of our approach across different domains, we additionally use three evaluation datasets for non-target evaluation, denoted as Eval-Neg.
First, we remove English from the VoxLingua107~\cite{valk2021slt} dataset and use the leftover 106 languages for our out-domain evaluation denoted as VL106. We randomly sample 200 utterances from each of the 106 Languages, resulting in 21,200 utterances split into 209,240 one-second samples.
Second, we use the Musan~\cite{snyder2015musan} dataset to test the noise-robustness of our approach. 
\begin{table}[htbp]
    \centering
    \begin{tabular}{ll|ll}
        \toprule
        Dataset & Task & Purpose & \# Samples \\
        \midrule
        GSCV1 &  \multirow{5}{*}{KWS} & Train & 51,088 (18 h) \\
        GSCV1 &  & Eval & 6,835 (1.9 h) \\
        \cline{3-4}
        GSCV2 & & Eval-Neg (In-domain) & 67,284 (18 h) \\
        Musan &  & Eval-Neg (Noise) &  394,432 (110 h) \\
        VL106 & & Eval-Neg (Out-domain) & 209,240 (68 h) \\
        \hline
        AS & \multirow{2}{*}{AT} & Train & 21,155 (58 h) \\
        AS &  & Eval & 18,229 (50 h) \\
        \bottomrule
    \end{tabular}
    \caption{Sample sizes and the duration for each respective dataset used in this work. All evaluation samples for KWS are of length 1s, while AT evaluation is done using 10s long samples.}
    \label{tab:dataset}
\end{table}
Musan consists of 394,432 samples with music, speech, and babble noises.
Third, we use the Google Speech Commands V2 (GSCV2)~\cite{warden2018speech} dataset as our in-domain non-target evaluation dataset, where we remove all 10 target keywords and obtain 67,284 individual test samples.
For all KWS evaluation datasets, we split all available audio clips into 1 s chunks, to match the training dataset duration.
The datasets are summarized in \Cref{tab:dataset}.

\subsection{Setup}

Our work focuses on using a common MobileNetV2~\cite{sandler2018mobilenetv2} back-end as our default neural network due to its small parameter size of 2.9 M.
        

Regarding front-end feature extraction, all audio clips are converted to a 16 kHz sampling rate. 
We use log-Mel spectrograms (LMS) with 64 bins extracted every 10 ms with a window of 32ms.
Batch-wise zero padding to the longest clip within a batch is applied for all experiments.
Training runs with a batch size of 64 for at most 500 epochs using Adam optimization~\cite{kingma2014adam} with a starting learning rate of 0.001.
During training, we use mean average precision (mAP) as our primary validation metric.
The top-4 models achieving the highest mAP on the joint held-out validation dataset (GSCV1-valid and AS-balanced) are submitted for evaluation.
The neural network back-end is implemented in Pytorch~\cite{PaszkePytorch}.

Assume that $\mathbf{x}$ is an input feature of some length, $\mathbf{y} \in [0,1]^{C+K}$ the corresponding label, $\mathcal{F}(\cdot)$ the trainable neural network, and $\hat{\mathbf{y}} \in [0,1]^{C+K}$ the model prediction, i.e., $\hat{\mathbf{y}} = \mathcal{F}(\mathbf{x})$. 
UniKW-AT is optimized using the binary cross entropy (BCE) objective defined as:
\begin{equation}
    \label{eq:bce}
    \mathcal{L}_{\text{BCE}}(\mathbf{\hat{y}}, \mathbf{y}) = \mathbf{y} \log{\hat{\mathbf{y}}} + (\mathbf{1}-\mathbf{y}) \log(\mathbf{1} - \hat{\mathbf{y}}),
\end{equation}
where $\mathbf{y}$ can have multiple positive labels.

One of the core difficulties in jointly modeling KWS and AT are the differences in their training sample duration.
In our experiments, the GSCV1 dataset contains only samples of duration less than $t=1$ s, while the majority of training samples in AS are of length 10 s.
A simple merge of both datasets during training would lead to a large amount of padding (for the GSCV1 dataset).
We provide an ablation study using the na\"ive data merging approach in \Cref{ssec:ablation_duration}.

\subsection{Metrics}
\label{ssec:metrics}

For KWS, we use accuracy as the default evaluation metric, while AT is mainly evaluated using mAP.
Computing accuracy for the UniKW-AT approach requires the use of post-processing since the model can predict multiple labels at once.
The core problem lies in the label ambiguity of speech-related labels since these are present for non-target and target keywords.
For example, when presented with a target keyword, our approach will likely predict the tuple (speech, target-keyword), but in the case of a nontarget-keyword, our approach will only predict speech.
Thus we compute the single-target output label as:
\begin{equation}
\label{eq:accuracy_bce}
    \bar{y} = \begin{cases}
    \argmax \mathbf{\hat{y}}_{\text{KWS}}, & \text{if } \max \mathbf{\hat{y}}_{\text{KWS}} \geq \gamma \\
    \argmax \mathbf{\hat{y}_{\text{AT}}},              & \text{otherwise}
\end{cases}
\end{equation}
where $\hat{\mathbf{y}} = \left[\mathbf{\hat{y}}_{\text{AT}}, \mathbf{\hat{y}}_{\text{KWS}} \right] \in [0,1]^{C+K} $ is the raw model prediction vector, $\bar{y}$ the predicted label index and $\gamma$ a decision threshold.
We optimized this threshold on the GSCV1 held-out validation dataset and obtained $\gamma = 0.4$ as our default choice.
Finally, when evaluating on AS, we only output $\hat{\mathbf{y}}_{\text{AT}} \in [0,1]^{C}$ as the representative prediction vector.

\section{Results}
\label{sec:results}

\subsection{UniKW-AT}
\label{ssec:multi_label_vs_single}

The results of our proposed UniKW-AT approach can be seen in \Cref{tab:refined_labels}.
UniKW-AT achieves a comparable performance to other methods in regards to both AT and KWS.
Notably, it outperforms the EfficientNet-B2 (15.70\%), AST (14.80\%), ResNet-50 (31.80\%), and CNN14 (27.80\%) methods on the balanced AS training set in regards to mAP performance (33.42\%) without additional data augmentation or external pretraining.
Further, our UniKW-AT method also contains the fewest parameters (2.9 M) within all investigated AT models.

Moreover, we observe that our approach only marginally lacks behind other models in terms of GSCV1 performance, achieving an accuracy of 97.53\%, without further data augmentation.
The much larger Wav2KWS~\cite{seo2021wav2kws} is the only approach to outperform UniKW-AT with an accuracy of 97.90\%, while also containing $\approx$ 80 times more parameters.
In addition, due to the benefits of the previously stated BCE training regime (see \Cref{sec:method}), we can remove the output layer's weights and biases, stripping down the model size to 2.2M, denoted as ``Ours-Strip''.
\begin{table}[htbp]
    \centering
    \begin{tabular}{ll|rr}
    \toprule
    Approach & \#Params (M) & GSCV1  & AS \\
    \midrule
    TC-ResNet8$^{\star}$~\cite{choi19_interspeech} &  0.06 & 96.10 & - \\
    TC-ResNet14-1.5$^{\star}$~\cite{choi19_interspeech} &  0.3 & 96.60 & - \\
    NAS1~\cite{Mo2020} & 0.28 & 97.06 & - \\
    NAS2~\cite{Mo2020} & 0.88 & 97.22 & - \\
    KWT-2$^{\star}$~\cite{berg21_interspeech} & 2.4 & 97.36 & -  \\
    KWT-3$^{\star}$~\cite{berg21_interspeech} & 5.3 & 97.24 & -  \\
    Wav2KWS$^{\star}$~\cite{seo2021wav2kws} & 225 & 97.90 & - \\
    \hline
    MobileNetV2$^{\dagger\star}$~\cite{gong2021psla} & 2.9 & - & 26.50 \\
    EfficientNet-B0$^{\dagger\star}$~\cite{gong2021psla} & 5.3 & - & 33.50 \\
    EfficientNet-B2~\cite{gong2021psla} & 13.6 & - & 15.70 \\
    EfficientNet-B2$^{\dagger\star}$~\cite{gong2021psla} & 13.6 & - & 34.06 \\
    ResNet-50$^{\dagger\star}$~\cite{gong2021psla} & 25.6 & - & 31.80 \\
    CNN14$^{\star}$~\cite{Kong2020d} & 76 & - & 27.80 \\
    AST~\cite{gong21b_interspeech} & 87 & - & 14.80\\
    \hline
    Ours-Strip & 2.2 & 97.53 & - \\
    Ours & 2.9 & 97.53 & 33.42 \\
    
        \bottomrule
    \end{tabular}
    \caption{A comparison between our proposed joint AT-KWS approach against other works in literature. Results for GSCV1 use accuracy, while AS use mAP. Approaches denoted with $^{\dagger}$ use pretraining on Imagenet and $^{\star}$ denotes additional data augmentation. Results with ``-'' means not available. ``Ours-Strip'' represents the removal of all AS-related output nodes.}
    \label{tab:refined_labels}
\end{table}
We conclude that our results indicate that UniKW-AT provides excellent performance for both individual tasks.

\subsection{Noise robustness}
\label{ssec:noise_robustness}

One of the benefits of the UniKW-AT approach is that it naturally enhances the noise robustness compared to standard KWS models.
A core reason for this is the addition of AS into the training regime.
Thus, we compare our proposed approach to a na\"ive CE trained baseline with the addition of AS as a noise source.
\begin{table}[htbp]
    \centering
    \begin{tabular}{l|rr|r}
    \toprule
    \multirow{1}{*}{Dataset} & Baseline & Baseline w/AS & Ours\\
        \midrule
    GSCV1 &  97.63 & \textbf{97.75} & 97.53  \\
    \hline
    GSCV2 & 97.47 & \textbf{98.90} & \textbf{98.90}  \\
    VL106 & 83.48 & 99.81 & \textbf{99.91}  \\
    Musan & 84.88 & 99.80 & \textbf{99.86} \\
        \bottomrule
    \end{tabular}
    \caption{KWS performance comparison regarding noise robustness. ``Baseline'' refers to using GSCV1 as the training dataset, while `Baseline w/AS`'' utilizes Audioset as additional non-target data. Best results in bold.}
    \label{tab:noise_robustness}
\end{table}
The results regarding noise-robustness are provided in \Cref{tab:noise_robustness}.
It can be seen that the na\"ive baseline performs well on in-domain evaluation, achieving an accuracy of 97.63\% and 97.47\% on the GSCV1 and GSCV2 evaluation sets, respectively.
However, a significant performance drop is seen when evaluating the na\"ive baseline in out-domain (VL106) and noise (Musan) scenarios, since the baseline has been trained on clean data.
When adding AS to the baseline training regime, performance steadily improves on all datasets compared to the baseline.
With additional AS data the baseline approach slightly outperforms UniKW-AT on the clean GSCV1 evaluation set, achieving an accuracy of 97.75\% against UniKW-AT's 97.53\%.
On the other hand, the non-target keyword evaluation on all reported datasets (GSCV2, VL106, Musan) shows that our UniKW-AT method outperforms the baseline.

\subsection{Real-world evaluation}
\label{sec:real_world_eval}

In order to verify the effectiveness of the proposed approach, we explore its real-world usability.
Here we use a private in-house dataset containing samples for our voice assistant ``XiaoAi''.
We collect 131 h of real-world training samples, split into 144,582 individual utterances with a length $t$ of 3 s.
Within these 144,582 training samples, 100,582 (70\%) contain the target keyword ``XiaoAi Tongxue''.
We evaluate our findings on two distinct datasets according to our evaluation framework (see \Cref{ssec:metrics}).
First, we use our ``Common'' dataset, containing 93,800 test utterances (80 h) with a balanced split of 50\% positives (``XiaoAi Tongxue'') and negative samples.
Second, we use a difficult non-target evaluation dataset, denoted as ``Difficult'', which contains 3,731 (3 h) samples. 
The Difficult evaluation set consists of non-target keyword samples similar to the target keywords e.g., ``Zhewei Tongxue'' and other cases of false-alarms such as child babbling and vocalized music.
Here we compare UniKW-AT against an in-house KWS model, denoted as ``Base-Inhouse''.
\begin{table}[htbp]
    \centering
    \begin{tabular}{l|rr}
    \toprule
    Dataset & Base-Inhouse & UniKW-AT (Ours) \\
        \midrule
    Common &  \textbf{99.98} & {99.85} \\
    Difficult & 0 & \textbf{82.73}  \\
        \bottomrule
    \end{tabular}
    \caption{Evaluation on our real-world in-house datasets. ``Base-Inhouse'' is a company internal KWS model. Best results in bold.}
    \label{tab:real_world_xiaoai}
\end{table}
The results are displayed in \Cref{tab:real_world_xiaoai}.
Real-world evaluation results show a similar trend to those in \Cref{tab:noise_robustness}.
The proposed UniKW-AT approach sightly underperforms on the Common evaluation set (99.85\% vs. 99.98\%), but significantly outperforms the baseline on the Difficult dataset (0\% vs. 82.73\%).
These results show that UniKW-AT can enhance noise robustness in real-world KWS applications.
Further performance gains can be achieved by tuning the decision threshold $\gamma$.
For example, setting $\gamma=0.01$, one obtains 99.96\% and 57.45\% on the Common and Difficult test sets, respectively.

\subsection{Ablations}
\label{ssec:ablation_duration}

In this section we explore the impact of two crucial components of our work:
The PSL module and the effect of random cropping on the AT branch.
The removal of random cropping, denoted as ``$-$ Crop'' leads to zero padding samples from the KWS dataset to match the Audioset's length i.e., 10 s.
The removal of PSL uses the original hard labels provided in Audioset.
The results of these ablation studies are displayed in \Cref{tab:ablations}.

\begin{table}[htbp]
    \centering
    \begin{tabular}{l|rrrr|r}
    \toprule
    Method & GSCV1 & GSCV2 & VL106 & Musan & AS\\
        \midrule
    Ours & {97.53} & 98.92 & 99.91 & {99.86} & {33.42}  \\
    \cline{2-6}
    $-$ Crop & 69.59 & 99.88 & 99.97 & 99.96 & 17.37 \\
    $-$ PSL & 96.94 & 98.23 & 99.89 & 99.85 & 20.77 \\
        \bottomrule
    \end{tabular}
    \caption{Ablation studies of removing random cropping (10 s training) and the removal of PSL. GSCV1/2, VL106 and Musan use accuracy as their metric, while AS uses mAP. }
    \label{tab:ablations}
\end{table}

\paragraph*{Impact of random cropping}
It can be seen that random cropping of the AT branch has a profound impact on the KWS performance.
Without it, the GSCV1 performance significantly degrades from 97.53\% to 69.59\% accuracy.
Unsurprisingly, the performance on the three negative evaluation datasets improves, since the model rarely detects the presence of a keyword.
Moreover, removing random cropping also results in the worst performance of 17.37\% mAP on the AS evaluation set.

\paragraph*{Impact of PSL}

In line with the initial work for PSL~\cite{dinkel2022_icassp}, the performance on AS significantly improves (20.77\% to 33.42\%) by re-labeling the balanced AS training set.
The results show that without PSL, performance uniformly degrades across all evaluation datasets.
Surprisingly, the performance on the GSCV1 dataset improves with PSL from 96.94\% to 97.53\%.

\section{Discussion}
\label{sec:discussion}

We would emphasize that our work is practically focused, such that we prefer lightweight models with an adequate performance over large models with excellent performance.
Future work could focus on using more optimal models for this task since the proposed MobileNetV2 could be considered excessively large within the KWS community.
Incidentally, compared to results within the AT community, our proposed MobileNetV2 framework is considered lightweight.

Even though this work indicates that UniKW-AT only marginally improves performance against common KWS baselines, it should be emphasized that UniKW-AT is a much more flexible framework for real-world KWS.
One potential usage is to devise different post-processing approaches for different environments.
For example, conditional KWS can be easily implemented i.e., wake-up only if children are speaking.
Most importantly, we believe that UniKW-AT's additional use case in form of AT would be justification enough to replace a pre-existing KWS model with UniKW-AT.
Finally, it is noteworthy that UniKW-AT outperforms the commonly used YAMNet~\cite{ellis_yamnet} model (30.6 vs. 33.4 mAP) for AT tasks with fewer parameters (3.7 M vs. 2.9 M).

\section{Conclusions}
\label{sec:conclusion}

This work proposes a novel approach to jointly modeling KWS and AT, denoted as UniKW-AT.
Our approach models these two tasks as an extension of AT by using a single multi-label BCE training objective with a lightweight MobileNetV2 back-bone.
UniKW-AT achieves an accuracy of 97.53\% on the common GSCV1 benchmark dataset and an mAP of 33.4 on the Audioset evaluation subset, outperforming other works explicitly optimized for each respective task.
Further, we provide results regarding noise-robustness using three out-of-domain scenarios and a study for real-world evaluation.
For all three out-of-domain evaluation datasets, UniKW-AT shows superior performance to standard KWS models, rejecting up to 99.86\% of noisy samples in the Musan dataset and 99.86\% of samples in the VoxLingua106 dataset.
Further, UniKW-AT also significantly enhances noise robustness against difficult samples in our real-world ``XiaoAi'' dataset, rejecting up to 82.73\% of non-target samples.
We believe that our work could spark future research for real-world applications of audio tagging and keyword spotting.

\newpage
\bibliographystyle{IEEEtran}

\bibliography{bib}


\end{document}